\documentclass[
  aps,
  pre,
  reprint,
  groupedaddress,
  longbibliography,
  amsmath,
  amssymb
]{revtex4-2}

\usepackage{graphicx}
\usepackage{newtxtext}
\usepackage{newtxmath}
\usepackage{bm}

\usepackage[
  colorlinks=true,
  linkcolor=blue,
  citecolor=blue,
  urlcolor=blue
]{hyperref}

\begin{document}

\title{
  Moment hierarchy and exact steady-state solutions in a Callaway lattice Boltzmann model for phonon hydrodynamics
  }

\author{Shimpei Saito}
\email[Contact author: ]{s.saito@aist.go.jp}
\affiliation{
  National Institute of Advanced Industrial Science and Technology (AIST), 
  1-2-1 Namiki, Tsukuba 305-8564, Japan
}

\date{\today}

\begin{abstract}
We analyze the moment hierarchy of a Callaway lattice Boltzmann model for phonon hydrodynamics and derive exact steady-state solutions for a driven planar channel with diffuse walls. 
The collision operator decomposes the population space into conserved, heat-flux, and fast kinetic sectors, whose coupling through streaming determines the transport response. 
For the two-dimensional, eight-velocity (D2Q8) lattice Boltzmann model, we derive a fully discrete closure that relates the heat flux to transverse kinetic moments.
The higher-order transverse kinetic moment contains both a second-difference term and an explicit contribution from the imposed energy drop. 
Diffuse-wall population constraints then determine the heat-flux and kinetic-moment profiles without fitted parameters. 
The resulting solution reproduces the numerical bulk moment amplitudes and wall slip, while an exact wall relation separates the contribution of the higher-order moment from deviations from the leading gradient approximation.
Finite-wave-number analysis further resolves the collision-sector mixing associated with longitudinal propagation. 
These results provide a unified description of collision relaxation, the moment hierarchy, and boundary response within the specified discrete model.
\end{abstract}

\maketitle

\section{Introduction}
\label{sec:introduction}

Phonons, the elementary excitations of lattice vibrations, can transport heat collectively in a manner analogous to fluid flow.
Such hydrodynamic transport can emerge when normal phonon scattering, which conserves crystal momentum, is much faster than resistive scattering, which relaxes it.
Characteristic phenomena include second sound and phonon Poiseuille flow~\citep{Callaway1959, GuyerKrumhansl1966Solution, GuyerKrumhansl1966Thermal, Hardy1970, HardyAlbers1974}.
Collision-spectrum approaches have been used in phonon kinetic theory, ranging from eigenmode solutions of the linearized Boltzmann transport equation to modern relaxon descriptions of the full scattering matrix~\citep{Chaput2013,CepellottiMarzari2016, SimoncelliMarzariCepellotti2020}. 
First-principles calculations have identified hydrodynamic phonon transport in graphene~\citep{CepellottiEtAl2015, LeeBroidoEsfarjaniChen2015}.
Subsequent experiments have reported second sound and phonon Poiseuille flow in graphite~\citep{HubermanEtAl2019, DingEtAl2022SecondSound, HuangEtAl2023}.
Evidence for phonon Poiseuille flow has also been reported in black phosphorus~\citep{Machida2018Poiseuille} and sapphire~\citep{Kawabata2025Sapphire}.
Other studies have explored vortical and backflow responses, further highlighting the spatial structure of hydrodynamic heat transport~\citep{GuoEtAl2021Vortex, DiLucenteLibbiMarzari2026, DragasevicRajkovSimoncelli2026}.

Finite discrete-velocity lattice Boltzmann formulations provide a compact mesoscopic description of phonon hydrodynamics~\citep{Guyer1994, JiaungHo2008, NabovatiSellanAmon2011, GuoWang2016}.
In particular, \citet{GuoWang2022} developed a no-rest two-dimensional, eight-velocity (D2Q8) Callaway scheme by rewriting the dual-relaxation kinetic equation as an effective single-relaxation collision term with a resistive source.
Their model recovered the phonon hydrodynamic equation, captured the transition from diffusive to hydrodynamic transport, reproduced Poiseuille flow and second sound, and incorporated diffuse-reflection boundary conditions.
Extending the phonon lattice Boltzmann method (LBM) beyond the near-continuum regime requires careful treatment of angular resolution and propagation anisotropy.
\citet{HammerFritzBedoyaMartinez2021} addressed these issues using an enlarged set of propagation directions.
More recently, \citet{LiEtAl2026} combined multi-directional propagation with a uniform assignment of base lattice directions and demonstrated improved predictions across quasi-ballistic, hydrodynamic, and diffusive benchmark cases.
These developments also motivate a complementary analytical question: how do collision relaxation, moment coupling, and population boundary conditions determine the transport response of a given discrete model?

The reduced dynamics considered in this study starts from the Callaway-type kinetic equation~\citep{Callaway1959, JiaungHo2008, LeeSihnRoyFarmer2012}
\begin{equation}
\partial_t f_i
+
\bm{c}_i\cdot\nabla f_i
=
-\frac{f_i-f_i^{N}}{\tau_N}
-\frac{f_i-f_i^{R}}{\tau_R},
\label{eq:callaway_kinetic}
\end{equation}
where $f_i^N$ and $f_i^R$ denote the normal and resistive equilibria, respectively, and $\tau_N$ and $\tau_R$ are the corresponding relaxation times.
Our main result is a fully discrete steady closure that predicts the heat flux, transverse kinetic moments, and wall slip without fitted coefficients. 

We organize the analysis around conserved, heat-flux, and fast kinetic eigenspaces while retaining the raw Cartesian moments to preserve the physical interpretation of the transport couplings. 
Moment-space collision has long been used in multiple-relaxation-time (MRT) LBMs~\citep{dHumieres1992, LallemandLuo2000, dHumieresEtAl2002, 
LallemandLuo2003, Adhikari2008, KaehlerWagner2013}; 
in the present model, the relaxation rates are fixed by the reduced Callaway operator. 
The multiplicities of the eigenspaces follow directly from the nested equilibrium projectors. 
We then examine which moments in these eigenspaces are coupled by streaming and boundary conditions, and how these couplings determine the transport response. 

Exact solutions of moment hierarchies with population boundary conditions have been obtained for lattice Boltzmann models.
\citet{Ansumali2007ExactHierarchy} obtained exact solutions for planar Couette flow.
\citet{KimPitsch2008} derived an exact slip-velocity solution for finite-Knudsen-number Poiseuille flow in a higher-order lattice Boltzmann model and examined the influence of higher-order moment relaxation times.
For phonon transport, Guo and Wang~\cite{GuoWang2022} obtained a heat-flux solution from continuum phonon hydrodynamics.
Here, we solve the steady boundary-value problem specified by the Callaway lattice update, periodic energy-drop driving, and diffuse-wall population reconstruction.
The resulting fully discrete solution determines the heat flux and the transverse kinetic-moment amplitudes without fitted parameters, and quantifies finite-width corrections to the leading bulk and wall closures.
A longitudinal spectral comparison then connects these steady-state results to the conserved, heat-flux, and fast kinetic sectors of the same collision operator.

\section{Discrete phonon model and moment hierarchy}
\label{sec:model}

\subsection{Fully discrete lattice evolution}
\label{sec:bte}

Using a unit time step, we advance the kinetic equation in Eq.~(\ref{eq:callaway_kinetic}) with the lattice update
\begin{equation}
f_i(\bm{x}+\bm{c}_i,t+1)-f_i(\bm{x},t)
=
-\frac{f_i-f_i^{N}}{\tau_N}
-\frac{f_i-f_i^{R}}{\tau_R}.
\label{eq:lbe_update}
\end{equation}
Here, $f_i=f_i(\bm{x},t)$ denotes the population associated with the discrete velocity $\bm{c}_i$. 
Equation~(\ref{eq:lbe_update}) makes explicit the collision--streaming dynamics whose eigenspace content is analyzed below.

The model energy density and heat flux are defined as the zeroth and first velocity moments, respectively:
\begin{equation}
E=\sum_i f_i,
\qquad
\bm{q}=\sum_i \bm{c}_i f_i.
\label{eq:hydrodynamic_moments}
\end{equation}
The normal process conserves both $E$ and $\bm{q}$, whereas the resistive process conserves $E$ but relaxes $\bm{q}$~\citep{Callaway1959, JiaungHo2008}.

The direct lattice relaxation parameters are denoted by $\tau_N$ and $\tau_R$. 
Their relation to the physical scattering times in the Guo--Wang formulation includes half-time-step corrections, as given in Sec.~\ref{sec:guowang_equivalence}. 
Normal scattering therefore controls the relaxation of higher-order kinetic moments toward a displaced local equilibrium, whereas resistive scattering determines the decay of the slow heat-flux sector.

\subsection{D2Q8 discrete-velocity model}
\label{sec:d2q8}

We consider the D2Q8 velocity set without a rest population
\begin{equation}
\begin{aligned}
(\bm{c}_1,\ldots,\bm{c}_8)
={}\bigl[&(1,0),(0,1),(-1,0),(0,-1),\\
& (1,1),(-1,1),(-1,-1),(1,-1)\bigr].
\end{aligned}
\label{eq:d2q8}
\end{equation}
We use lattice units with
\begin{equation}
\Delta x=\Delta t=1,
\qquad
c
=
\frac{\Delta x}{\Delta t}
=
1.
\end{equation}
Here, $c$ is the lattice propagation speed.
In the isotropic gray Debye interpretation proposed by \citet{GuoWang2022}, it is related to the phonon group speed $v_g$ by $c=\sqrt{3/5}\,v_g$, as determined by matching the hydrodynamic transport coefficients.
Thus, $c=1$ corresponds to $v_g=\sqrt{5/3}$ in the present lattice units.
D2Q8 is used as a compact no-rest realization of the reduced Callaway dynamics. 
Its limited velocity set does not provide an exact angular representation of the phonon Boltzmann equation~\citep{NabovatiSellanAmon2011, HammerFritzBedoyaMartinez2021}.
The steady solutions derived below are exact for the specified lattice update and boundary reconstruction under the stated parameter conditions; their derivation does not invoke a near-continuum approximation.
Hence, their accuracy as approximations to continuous-angle phonon transport is a separate question that requires an assessment of velocity-space discretization, particularly for wall observables and transport beyond the near-continuum regime.

Consistent with previous Callaway-type phonon LBM formulations~\citep{JiaungHo2008, LeeSihnRoyFarmer2012,GuoWang2022}, we use the following discrete resistive and normal equilibria:
\begin{equation}
f_i^{R}=w_iE,
\qquad
f_i^{N}=w_iE+\lambda_i\,\bm{c}_i\cdot\bm{q},
\label{eq:population_equilibria}
\end{equation}
with
\begin{equation}
w_i=
\begin{cases}
2/9, & i=1,\ldots,4,\\
1/36, & i=5,\ldots,8,
\end{cases}
\qquad
\lambda_i=
\begin{cases}
1/3, & i=1,\ldots,4,\\
1/12, & i=5,\ldots,8.
\end{cases}
\label{eq:d2q8_equilibrium_coefficients}
\end{equation}
These population-space equilibria are equivalent to the moment-space forms introduced in the next section.

\subsection{Cartesian moment hierarchy}
\label{sec:moments}

For a discrete distribution $f_i$, we define the raw Cartesian moments as
\begin{equation}
m_{pq}
=
\sum_{i=1}^{8}
c_{ix}^{p}c_{iy}^{q}f_i.
\label{eq:rawmoment}
\end{equation}
Rather than introducing linear combinations at the moment definition level, we retain the unmixed Cartesian components
\begin{equation}
\bm{m}
=
(
m_{00},
m_{10},m_{01},
m_{20},m_{02},m_{11},
m_{21},m_{12}
)^{\mathrm T}.
\label{eq:momentvector}
\end{equation}

This choice provides a direct hierarchy according to tensorial order: 
$m_{00}$ represents the energy density;
$m_{10}$ and $m_{01}$ are the two components of the heat flux;
$m_{20}$, $m_{02}$, and $m_{11}$ form the second-order tensor sector;
and $m_{21}$ and $m_{12}$ constitute the remaining independent higher-order moments. 
The eight-dimensional moment set is thus complete for the D2Q8 population space.

The discrete velocities satisfy
$c_{\alpha}^{3}=c_{\alpha}$ and
$c_{\alpha}^{4}=c_{\alpha}^{2}$;
hence, higher polynomial moments are not all independent. 
For example,
\begin{equation}
m_{22}
=
m_{20}+m_{02}-m_{00},
\label{eq:m22alias}
\end{equation}
for the present no-rest D2Q8 lattice. 
The representation therefore closes naturally without introducing an additional fourth-order degree of freedom.

\section{Moment-based collision and relaxation spectrum}
\label{sec:collision}

\subsection{Callaway projectors and collision spectrum}
\label{sec:general_spectrum}

Let $P_R$ and $P_N$ denote the linear projectors that map a population state $\bm f$ onto the resistive and normal equilibria, respectively, with $P_R\bm f=\bm f^{R}$ and $P_N\bm f=\bm f^{N}$.
The collision part of Eq.~(\ref{eq:lbe_update}) can be written in population space as
\begin{equation}
C
=
-\alpha(I-P_N)-\beta(I-P_R),
\label{eq:general_callaway_operator}
\end{equation}
where $I$ denotes the identity operator in population space, $\alpha=\tau_N^{-1}$, and $\beta=\tau_R^{-1}$.

For the reduced class considered here, the resistive equilibrium retains only the energy density, whereas the normal equilibrium retains the energy and the $d$ heat-flux components. 
For these equilibrium maps, conservation of the retained moments gives $P_RP_N=P_NP_R=P_R$.
The resistive-equilibrium subspace is contained in the normal-equilibrium subspace. 
The corresponding projector ranks are $\operatorname{rank}(P_R)=1$ and $\operatorname{rank}(P_N)=d+1$.

The nested projector structure separates the population space into three invariant sectors. 
We define
\begin{equation}
P_E\equiv P_R,
\qquad
P_q\equiv P_N-P_R,
\qquad
P_k\equiv I-P_N,
\label{eq:sector_projectors}
\end{equation}
which project onto the conserved energy sector, additional heat-flux sector retained by the normal equilibrium, and remaining kinetic sector, respectively. 
As the equilibrium subspaces are nested, these projectors are mutually annihilating and satisfy $P_E+P_q+P_k=I$.

With $\gamma=\alpha+\beta$, the collision operator becomes
\begin{equation}
C=-\beta P_q-\gamma P_k,
\label{eq:general_callaway_split}
\end{equation}
so that
\begin{equation*}
CP_E=0,
\qquad
CP_q=-\beta P_q,
\qquad
CP_k=-\gamma P_k.
\end{equation*}
Thus, the collision space decomposes into three sectors: one conserved energy mode, $d$ slow heat-flux modes, and $q-d-1$ fast kinetic modes. 
For D2Q8, the multiplicities are $1+2+5$. 
At fixed $d$, increasing $q$ while keeping the normal-equilibrium subspace unchanged therefore adds only fast kinetic degrees of freedom.

The multiplicities themselves follow directly from the projector ranks.
Representative multiplicities for several common velocity sets are listed in the Supplemental Material to illustrate this rank-based counting.
The following analysis instead focuses on how finite wave number, channel geometry, and diffuse boundary conditions select particular combinations of moments and determine their couplings.

\subsection{Cartesian moment coordinates and relaxation eigenmodes}
\label{sec:moment_eigenmodes}

The population and Cartesian moment representations are related by $\bm{m}=M\bm{f}$, where $M$ is constructed directly from the Cartesian monomials in Eq.~(\ref{eq:rawmoment}). 
For the velocity ordering in Eq.~(\ref{eq:d2q8}), the transformation matrix is
\begin{equation}
M=
\begin{pmatrix}
1&1&1&1&1&1&1&1\\
1&0&-1&0&1&-1&-1&1\\
0&1&0&-1&1&1&-1&-1\\
1&0&1&0&1&1&1&1\\
0&1&0&1&1&1&1&1\\
0&0&0&0&1&-1&1&-1\\
0&0&0&0&1&1&-1&-1\\
0&0&0&0&1&-1&-1&1
\end{pmatrix}.
\label{eq:moment_matrix}
\end{equation}

The collision matrix is $C_m=MCM^{-1}$, and the local update is $\bm m^*=(I+C_m)\bm m$. 
If $R$ contains right eigenvectors of $C_m$, then $\bm a=R^{-1}\bm m$ obeys $\bm a^*=(I+\Lambda)\bm a$, where $C_mR=R\Lambda$.
The raw moments specify transport coordinates, whereas the relaxation amplitudes isolate the local decay channels. 
A homogeneous-relaxation example is provided in the Supplemental Material.

\subsection{D2Q8 Cartesian realization}
\label{sec:collision_spectrum_matrix}

With the rates defined above, the slow--fast separation may be measured by
\begin{equation}
\frac{\beta}{\gamma}
=
\frac{\tau_N}{\tau_N+\tau_R},
\label{eq:scale_separation}
\end{equation}
which tends to zero as the hydrodynamic separation $\tau_R/\tau_N$ increases.
The D2Q8 resistive and normal equilibria in Cartesian moments are
\begin{align}
\bm{m}^{R}
&=
\left(
E,0,0,\chi E,\chi E,0,0,0
\right)^{\mathrm T},
\\
\bm{m}^{N}
&=
\left(
E,q_x,q_y,\chi E,\chi E,0,\eta q_y,\eta q_x
\right)^{\mathrm T},
\end{align}
with $\chi=5/9$ and $\eta=1/3$.
The resulting collision matrix in the Cartesian basis is
\begin{equation}
C_m=
\begin{pmatrix}
0&0&0&0&0&0&0&0\\
0&-\beta&0&0&0&0&0&0\\
0&0&-\beta&0&0&0&0&0\\
\chi\gamma&0&0&-\gamma&0&0&0&0\\
\chi\gamma&0&0&0&-\gamma&0&0&0\\
0&0&0&0&0&-\gamma&0&0\\
0&0&\eta\alpha&0&0&0&-\gamma&0\\
0&\eta\alpha&0&0&0&0&0&-\gamma
\end{pmatrix}.
\label{eq:Cm}
\end{equation}

A convenient set of relaxation amplitudes is
\begin{equation}
\bm{a}
=\begin{pmatrix}
a_1\\a_2\\a_3\\a_4\\a_5\\a_6\\a_7\\a_8
\end{pmatrix}
=
\begin{pmatrix}
m_{00}\\
m_{10}\\
m_{01}\\
m_{20}-\chi m_{00}\\
m_{02}-\chi m_{00}\\
m_{11}\\
m_{21}-\eta m_{01}\\
m_{12}-\eta m_{10}
\end{pmatrix}.
\label{eq:relaxation_amplitudes}
\end{equation}
For the streamwise heat flux $q_x=m_{10}$ considered below, the transverse kinetic amplitudes $a_6=m_{11}$ and $a_8=m_{12}-\eta m_{10}$ play a central role in the steady moment hierarchy and wall closure.
The collision step for this set is diagonal,
\begin{equation}
a_j^{*}=
\begin{cases}
a_j, & j=1,\\
(1-\beta)a_j, & j=2,3,\\
(1-\gamma)a_j, & j=4,\ldots,8.
\end{cases}
\label{eq:a_collision}
\end{equation}
where the superscript $*$ denotes the post-collision state.
Thus, the two nonzero relaxation rates are fixed by the Callaway scattering times and not tuned as independent MRT parameters. 
The five-dimensional fast eigenspace is degenerate; hence, its internal basis is not unique.

\subsection{Relation to the Guo--Wang source formulation}
\label{sec:guowang_equivalence}

Relating the present formulation to that in Ref.~\citep{GuoWang2022} requires a parameter and flux conversion. 
In units of the time step, we denote the dimensionless physical resistive and combined scattering times by $T_R$ and $T_C$, with $T_C^{-1}=T_N^{-1}+T_R^{-1}$, and the lattice relaxation parameter by $\theta=T_C+1/2$. 
Let $\bm q^{\rm GW}$ be the source-corrected flux and $\bm q=\sum_i\bm c_i f_i$ the raw flux used here. 
The source correction and lattice step can be written as
\begin{align}
\bm q^{\rm GW}&=\frac{\bm q}{1+1/(2T_R)},\nonumber\\
f_i^*&=f_i-\frac{f_i-f_i^{\rm eq}}{\theta}+S_i,
\label{eq:gw_source_step}
\end{align}
where $f_i^{\rm eq}=w_iE+\lambda_i\bm c_i\cdot\bm q^{\rm GW}$ and
$S_i=-(1-1/(2\theta))\lambda_i\bm c_i\cdot\bm q^{\rm GW}/T_R$.
These are the source and flux conventions used by Ref.~\citep{GuoWang2022} expressed in the lattice units and weights of Eq.~(\ref{eq:d2q8_equilibrium_coefficients}).
Setting $s_R=1/T_R$, the raw flux can be expressed as
\begin{equation}
\bm q^*=\left[1-\frac{s_R}{1+s_R/2}\right]\bm q.
\end{equation}
Energy is unchanged, while the fast-sector component has zero energy and flux and is multiplied by $1-1/\theta$. 
Consequently, the one-step map is
\begin{equation}
B_{\rm GW}=P_E+(1-\beta)P_q+(1-\gamma)P_k,
\label{eq:guowang_equivalent_step}
\end{equation}
provided
\begin{equation}
\beta=\frac{1}{T_R+1/2},\qquad
\gamma=\frac{1}{T_C+1/2}.
\label{eq:gw_rate_mapping}
\end{equation}
The flux conversion is then $\bm q^{\rm GW}=(1-\beta/2)\bm q$.
Conversely, $T_R=\beta^{-1}-1/2$, $T_C=\gamma^{-1}-1/2$, and $T_N=(T_C^{-1}-T_R^{-1})^{-1}$. 
Positive physical times require $0<\beta<\gamma<2$ for finite $T_N$.
Thus, the direct and source forms have the same amplification matrix when this conversion and the same streaming operator are used. 
Equal numerical values of the scattering times in the two formulations do not, in general, imply equal updates. 
The model Knudsen numbers used below refer to the direct parameters $\tau_N=(\gamma-\beta)^{-1}$ and $\tau_R=\beta^{-1}$, not to $T_N,T_R$.

\section{Spectral connections to hydrodynamic transport}
\label{sec:analysis}

\subsection{Chapman--Enskog connection}
\label{sec:ce}

The hydrodynamic equation for the D2Q8 Callaway scheme has already been derived in detail by Guo and Wang \cite{GuoWang2022}. 
We retain only the coefficients needed below. 
A second-order expansion of the fully discrete collision--streaming dynamics gives
\begin{equation}
\partial_tE+
\left(1-\frac{\beta}{2}\right)\nabla\cdot\bm q
=
O(\varepsilon^3),
\end{equation}
and
\begin{equation}
\begin{aligned}
&\left(1-\frac{\beta}{2}\right)\partial_t\bm q
+\beta\bm q+\chi\nabla E \\
&\quad=
\eta\left(\frac{1}{\gamma}-\frac{1}{2}\right)
\left[
\nabla^2\bm q+\frac{1}{3}\nabla(\nabla\cdot\bm q)
\right]
+O(\varepsilon^3).
\end{aligned}
\label{eq:ce_q_final}
\end{equation}
With the conservative flux
\begin{equation}
\bm J=
\left(1-\frac{\beta}{2}\right)\bm q,
\end{equation}
the Chapman--Enskog thermal diffusivity $D_{\mathrm{CE}}$ and the squared transport lengths $\ell_{\mathrm{cont}}^2$ and $\ell_{\mathrm{CE}}^2$ are
\begin{align}
D_{\mathrm{CE}}
&=
\chi\left(\beta^{-1}-\frac12\right),
\\
\ell_{\mathrm{cont}}^2
&=
\frac{\eta}{\beta\gamma},
\label{eq:continuous_transport_length}\\
\ell_{\mathrm{CE}}^2
&=
\frac{\eta}{\beta}
\left(\gamma^{-1}-\frac12\right).
\label{eq:transport_lengths}
\end{align}
In the long-wavelength diffusive limit, $\bm J\simeq-D_{\mathrm{CE}}\nabla E$, so that $\partial_t E\simeq D_{\mathrm{CE}}\nabla^2 E$.
Here, $\ell_{\mathrm{cont}}$ follows from the leading-order closure of the continuous-time hierarchy, whereas $\ell_{\mathrm{CE}}$ includes the half-time-step correction of the fully discrete lattice.
Spectrally, $\beta$ controls the slow heat-flux sector, whereas $\gamma$ controls the fast kinetic correction responsible for nonlocal transport.

\subsection{Longitudinal moment closure}
\label{sec:longitudinal_closure}

For a one-dimensional perturbation with $\bm{k}=(k,0)$, the continuous-time kinetic equation closes exactly on
\begin{equation}
E=m_{00},
\qquad
q=m_{10},
\qquad
\pi=m_{20}-\chi E,
\end{equation}
with
\begin{align}
\partial_t E+\partial_x q &=0,
\\
\partial_t q+\chi\partial_x E+\partial_x\pi
&=-\beta q,
\\
\partial_t\pi+(1-\chi)\partial_x q
&=-\gamma\pi.
\label{eq:longitudinal_pi}
\end{align}
Thus, $\pi$ is the fast longitudinal kinetic amplitude. 
For a normal mode $\propto\exp(st+ikx)$,
\begin{equation}
s^3
+
(\beta+\gamma)s^2
+
(\beta\gamma+k^2)s
+
\chi\gamma k^2
=
0,
\label{eq:longitudinal_dispersion}
\end{equation}
which provides the continuous-time reference used below.

For temporal variations slow compared with the relaxation
rate $\gamma$,
\begin{equation}
\pi \simeq -\frac{1-\chi}{\gamma}\partial_x q.
\label{eq:longitudinal_slaving}
\end{equation}
The fast mode is therefore approximately determined by the gradient of the slow heat-flux mode.
The corresponding fully discrete Chapman--Enskog coefficient differs by the usual half-time-step correction,
$1/\gamma\rightarrow1/\gamma-1/2$.
This distinction is retained only when needed for comparison with the discrete amplification spectrum.

\subsection{Driven transverse hierarchy and exact steady closure}
\label{sec:transverse_hierarchy}

For channel transport, the walls lie on lattice nodes $y=0$ and $y=H$.
The streamwise direction is periodic with an imposed energy drop $\Delta E$ over $N_x$ lattice nodes; we use $N_x=3$ in the simulations.
During streaming, populations crossing the periodic boundary receive $+w_i\Delta E$ from the left and $-w_i\Delta E$ from the right.
We write $G=\Delta E/N_x>0$ for the imposed negative energy gradient.
After averaging over $x$, this driving contributes $w_i c_{ix}G$ to bulk streaming. 
The subsequent diffuse-wall reconstruction is applied separately. 
All the following moments are averaged over $x$ and evaluated after streaming and wall reconstruction.

The relevant transverse moments are
\begin{equation}
q=m_{10},\qquad p=m_{11}=a_6,\qquad r=m_{12}=\eta q+a_8.
\label{eq:transverse_hierarchy}
\end{equation}
Here, $q$ denotes $q_x(y)$ in the channel analysis. 
Although the velocity set is two-dimensional, the averaged steady fields depend only on $y$.
Transverse streaming selects the sequence $m_{10},m_{11},m_{12}$;
rotating the channel exchanges the roles of $a_8$ and $a_7=m_{21}-\eta q_y$. 
With $\sigma=1-\gamma$, their local collision step is
\begin{equation}
q^*=(1-\beta)q,\qquad p^*=\sigma p,\qquad
r^*=\sigma r+\eta\alpha q.
\label{eq:transverse_collision_exact}
\end{equation}
The horizontal population difference $j_0=f_1-f_3=q-r$ does not stream across $y$ and is not reconstructed at the walls. 
Its steady-state balance is $j_0=j_0^*+4G/9$. 
Substitution of Eq.~(\ref{eq:transverse_collision_exact}) therefore gives
\begin{equation}
a_8=\frac{2\beta}{3\gamma}q-\frac{4}{9\gamma}G.
\label{eq:a8_local_exact}
\end{equation}
This identity holds at both interior and wall nodes. 
Using a measured $q$ in this identity provides a diagnostic, whereas predicting $q$ requires solving for it from the driving and wall conditions.

Let us define $K=\eta+(1-\eta)\beta/\gamma$, $D_g=4/(9\gamma)$, and $L=K-\beta$, so that $r=Kq-D_gG$ and $r^*=Lq-\sigma D_gG$.
Eliminating $p$ from the steady streaming relations gives the scalar interior recurrence derived in Appendix~\ref{app:steady_closure},
\begin{equation}
q-\ell_D^2\Delta_2q=q_\infty,\qquad
q_\infty=\frac{\chi G}{\beta},
\label{eq:steady_scalar_recurrence}
\end{equation}
where $\Delta_2q_y=q_{y+1}-2q_y+q_{y-1}$ and
\begin{equation}
\ell_D^2=\frac{K(2-\gamma)-\beta}{2\beta\gamma}.
\label{eq:exact_discrete_length}
\end{equation}
The coefficient $\ell_D^2$ belongs to the exact steady difference equation. 
It is distinct from the continuum and Chapman--Enskog coefficients in Eqs.~(\ref{eq:continuous_transport_length}) and (\ref{eq:transport_lengths}).

Combining Eqs.~(\ref{eq:a8_local_exact}) and (\ref{eq:steady_scalar_recurrence}) yields the exact interior closure
\begin{equation}
a_8=B_2\Delta_2q-\frac{2}{27\gamma}G,\qquad
B_2=\frac{2\beta\ell_D^2}{3\gamma}.
\label{eq:a8_driven_closure}
\end{equation}
For a smooth profile, $\Delta_2q=\partial_y^2q + \partial_y^4q/12+\cdots$. 
Thus, $a_8$ contains a second-gradient contribution and an explicit driving contribution; a pure proportionality to $\partial_y^2q$ is not the general steady closure.

The leading continuous-time slaving relation remains
\begin{equation}
p\simeq-\frac{\eta}{\gamma}\partial_yq.
\label{eq:continuous_slaving}
\end{equation}
In a slowly varying Poiseuille ordering with $\partial_y=O(\varepsilon)$, $\beta=O(\varepsilon^2)$, $G/q=O(\varepsilon^2)$, and finite $\gamma$, the first-gradient response is $O(\varepsilon q)$ and $a_8=O(\varepsilon^2q)$. 
In particular, when $q_{\max}\propto GH^2$, the driving contribution in Eq.~(\ref{eq:a8_driven_closure}), normalized by $q_{\max}$, also scales as $H^{-2}$. 
A width exponent alone cannot distinguish the driving contribution from the second-gradient term. 
The following comparisons therefore test the amplitudes predicted by the full steady solution, not just their
power-law exponents.

\subsection{Exact steady profiles with diffuse walls}
\label{sec:exact_steady_profiles}

The on-node diffuse reconstruction sets $f_5=f_6$ at the bottom wall and $f_7=f_8$ at the top wall. 
Hence, $p_0=-r_0$ and $p_H=r_H$.
Together with Eq.~(\ref{eq:steady_scalar_recurrence}), these constraints fix the symmetric steady solution.

For the nonsingular cases considered here, $\ell_D^2>0$. 
We take $\ell_D$ to be its positive square root. 
We then define
\begin{equation}
\kappa=2\operatorname{arsinh}\!\left(\frac{1}{2\ell_D}\right),
\qquad
V=\frac{L\sinh\kappa}{1-\sigma\cosh\kappa}.
\label{eq:steady_spatial_rate}
\end{equation}
The symmetric solution is
\begin{align}
q_y&=q_\infty+A_c\cosh[\kappa(y-H/2)],\nonumber\\
a_{6,y}=p_y&=-VA_c\sinh[\kappa(y-H/2)],\nonumber\\
a_{8,y}&=(K-\eta)q_y-D_gG,
\label{eq:steady_analytic_profiles}
\end{align}
where the wall constraint determines
\begin{equation}
A_c=\frac{D_gG-Kq_\infty}
{K\cosh(\kappa H/2)+V\sinh(\kappa H/2)}.
\label{eq:steady_amplitude}
\end{equation}
Thus, $H$, $\tau_N$, $\tau_R$, and $G$ determine the heat flux and both transverse kinetic moments at every lattice node, including the walls. 
No numerical profile or fitted coefficient enters these expressions. 
The amplitude $A_c$ follows from the population boundary condition, rather than an imposed phenomenological slip length.
The spatial rate $\kappa$ retains the lattice difference operator; replacing it by $1/\ell_D$ would give a continuum approximation.
Appendix~\ref{app:steady_closure} supplies the derivation.

These profiles provide the parameter-only predictions tested in Figs.~\ref{fig:poiseuille_hierarchy} and \ref{fig:wall_kinetic}.
For the wall diagnostics, the analytical node values are evaluated with the same one-sided gradient estimator as the numerical data.
The solution is exact for the stated lattice update at steady state with the specified diffuse-wall reconstruction.

\subsection{Finite-wave-number amplification spectrum}
\label{sec:dispersion}

For a longitudinal perturbation $\delta\bm f\propto \exp(ikx)z^n$, the fully discrete dynamics is
\begin{equation}
\widehat{\bm f}^{\,n+1}
=
\mathcal A(k)\widehat{\bm f}^{\,n},
\qquad
z_a(k)=\exp[-i\omega_a(k)\Delta t].
\label{eq:zomega}
\end{equation}
The branch approaching $z=1$ at $k\to0$ has
\begin{equation}
-\ln|z_{\mathrm{diff}}(k)|
=
D_{\mathrm{CE}}k^2+O(k^4),
\end{equation}
providing a consistency check with the Chapman--Enskog limit. At finite $k$, a population-space amplification eigenvector $\bm v$ is projected onto the collision sectors.
We define
\begin{equation}
W_s(\bm v)=
\frac{\|P_s\bm v\|_2^2}
{\sum_{t\in\{E,q,k\}}\|P_t\bm v\|_2^2},
\quad s\in\{E,q,k\},
\label{eq:sector_weights}
\end{equation}
using the Euclidean norm of the eight population components. 
The weights sum to one and are independent of the scalar normalization of $\bm v$.
The projectors are mutually annihilating but need not be orthogonal in this norm, so the denominator is not assumed to equal $\|\bm v\|_2^2$.
These weights quantify projected population amplitudes in the stated representation.
They are not fractions of physical energy and are not invariant under an arbitrary change of coordinates or norm.

\section{Numerical results and physical interpretation}
\label{sec:results}

\subsection{Finite-$k$ hybridization and propagation}
\label{sec:crossover}

Using the amplification-spectrum construction discussed in Sec.~\ref{sec:dispersion}, the longitudinal branch that is diffusive when $k$ is small merges with another branch to form a propagating complex-conjugate pair at finite wave number.
We define the critical wave number $k_c$ by the onset of $\mathrm{Re}\,\omega\neq0$ and track the two longitudinal branches by
successive eigenvector overlap.

\begin{figure*}[tb]
\centering
\includegraphics[width=0.98\textwidth]{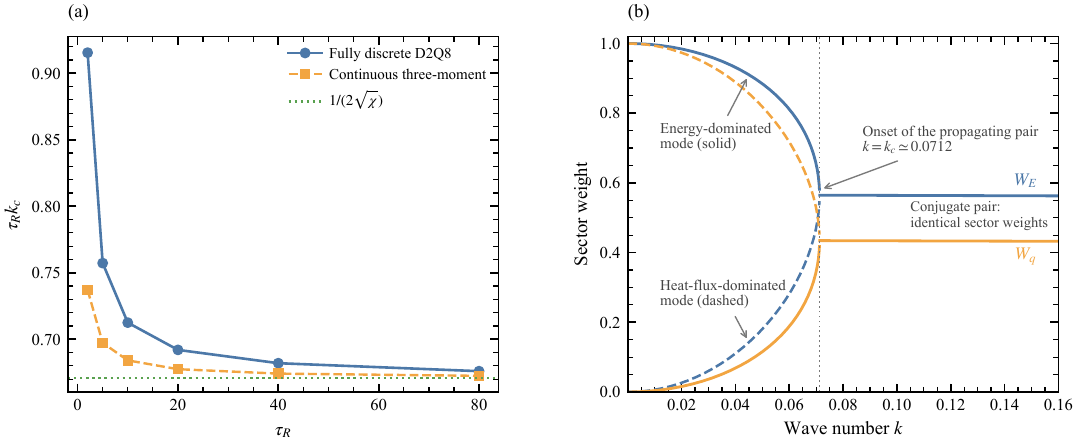}
\caption{
Finite-$k$ longitudinal spectrum.
(a) Scaled critical wave number $\tau_R k_c$ at fixed $\tau_N=1$ for the fully discrete D2Q8 model and continuous three-moment hierarchy; the dotted line denotes the two-moment asymptote $1/(2\sqrt{\chi})$.
(b) Collision-sector weights defined by Eq.~(\ref{eq:sector_weights}) for $\tau_N=1$ and $\tau_R=10$.
At small $k$, the two longitudinal branches are energy-dominated and heat-flux-dominated, respectively.
They become strongly mixed as $k$ approaches $k_c$ and, above $k_c$, form a complex-conjugate pair with identical sector weights; hence, only one curve is shown for each sector for $k>k_c$.
The omitted fast-sector weight $W_k$ remains below
$5\times10^{-3}$ over the range shown.
}
\label{fig:critical_wavenumber}
\end{figure*}

For reference, neglecting the fast longitudinal kinetic moment gives
\begin{equation}
s^2+\beta s+\chi k^2=0,
\end{equation}
and hence,
\begin{equation}
k_c^{(2)}
=
\frac{\beta}{2\sqrt{\chi}}
=
\frac{1}{2\sqrt{\chi}\tau_R}.
\label{eq:kc_two_moment}
\end{equation}
The fully discrete $k_c$ approaches both the continuous three-moment result and the two-moment asymptote as $\tau_R$ increases [Fig.~\ref{fig:critical_wavenumber}(a)].
The relative difference between the fully discrete and continuous three-moment values falls from approximately $24\%$ at $\tau_R=2$ to $0.5\%$ at $\tau_R=80$, while 
\begin{equation}
\tau_R k_c
\rightarrow
\frac{1}{2\sqrt{\chi}}
=
\frac{3}{2\sqrt5}
\simeq0.67082.
\label{eq:kc_asymptote}
\end{equation}

Figure~\ref{fig:critical_wavenumber}(b) resolves the eigenvector
reorganization directly. For $\tau_R=10$, initially energy-like and
heat-flux-like branches become strongly mixed near $k_c$ and form a
complex-conjugate pair above it. The fast-sector weight remains below $5\times10^{-3}$; hence, the crossover is predominantly a
conserved--slow hybridization.

\subsection{Phonon Poiseuille flow}
\label{sec:poiseuille}

We consider steady in-plane transport between diffusely reflecting
adiabatic walls and define
\begin{equation}
\mathrm{Kn}_{N,R}
=
\frac{c\tau_{N,R}}{H},
\label{eq:Kn}
\end{equation}
with $H$ as the channel width in lattice units and $c=1$. 
No mapping to a specific material is introduced.

\begin{figure*}[tb]
\centering
\includegraphics[width=0.88\textwidth]{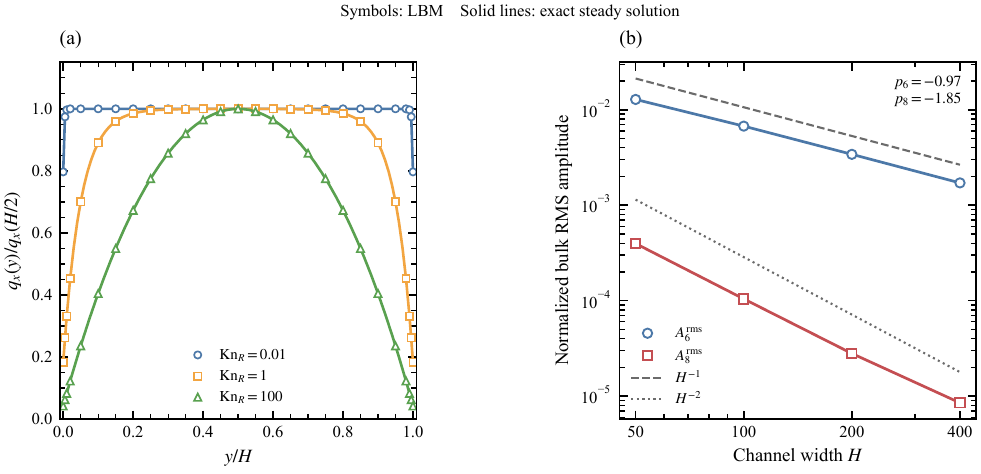}
\caption{
Poiseuille flow and transverse moment hierarchy.
Symbols denote LBM results, and solid lines denote predictions from the
exact steady discrete solution without fitted parameters.
(a) Normalized heat-flux profiles at fixed $\mathrm{Kn}_N=0.01$ and
$H=200$ for $\mathrm{Kn}_R=0.01$, $1$, and $100$.
Numerical and analytical profiles are each normalized by their own
centerline value; only selected numerical points are shown.
(b) Bulk RMS amplitudes $A_6^{\mathrm{rms}}$ and $A_8^{\mathrm{rms}}$
versus channel width at fixed $\tau_N=1$ and $\tau_R=10^5$.
Analytical amplitudes use the same bulk intervals as the numerical data
and are normalized by the analytical maximum heat flux.
Solid lines connect predictions at the sampled widths.
Dashed and dotted lines indicate $H^{-1}$ and $H^{-2}$ reference slopes,
respectively; $p_6$ and $p_8$ are fitted exponents of the numerical data.
}
\label{fig:poiseuille_hierarchy}
\end{figure*}

Figure~\ref{fig:poiseuille_hierarchy}(a) compares the numerical heat-flux profiles with the exact steady solution in Eq.~(\ref{eq:steady_analytic_profiles}).
We use $H=200$, $\tau_N=2$, and $\tau_R=2$, $200$, and $20000$, corresponding to
$\mathrm{Kn}_N=0.01$ and $\mathrm{Kn}_R=0.01$, $1$, and $100$.
Each numerical or analytical profile is normalized by its own centerline heat flux. 
The largest absolute difference between the normalized profiles is below $1.4\times10^{-10}$ over these three cases. 
The familiar transition to a Poiseuille-like profile therefore serves as a test of the exact discrete solution rather than a new transport phenomenon.

For the hierarchy comparison, we vary $H=50$, $100$, $200$, and $400$ at fixed $\tau_N=1$ and $\tau_R=10^5$. 
The model relaxation times remain unchanged, while the Knudsen numbers and normalized profile shape can change with $H$. 
We define
\begin{equation}
A_6^{\mathrm{rms}}=
\frac{\mathrm{rms}_{\rm bulk}(a_6)}{\max_y|q_x|},\qquad
A_8^{\mathrm{rms}}=
\frac{\mathrm{rms}_{\rm bulk}(a_8)}{\max_y|q_x|}.
\label{eq:bulk_rms_definition}
\end{equation}
The bulk intervals exclude $3$, $5$, $10$, and $20$ nodes from each end for the four widths, respectively. 
In zero-based coordinates they are $y=m,\ldots,H-m$, with $m$ the stated exclusion width. 
Numerical and analytical RMS amplitudes use identical intervals, and each is normalized by its own full-domain maximum heat flux.

The numerical amplitudes in Fig.~\ref{fig:poiseuille_hierarchy}(b) have fitted exponents
\begin{equation}
A_6^{\mathrm{rms}}\propto H^{-0.97},\qquad
A_8^{\mathrm{rms}}\propto H^{-1.85}
\label{eq:hierarchy_scaling}
\end{equation}
over the sampled widths. 
More stringently, the exact steady solution predicts their magnitudes without fitted coefficients: the largest relative differences from the numerical amplitudes are below $1.1\times10^{-10}$ for $A_6^{\mathrm{rms}}$ and $2.4\times10^{-8}$ for $A_8^{\mathrm{rms}}$. 
Thus, the agreement tests the driving-dependent closure, rather than only a nominal gradient
order. 
For comparison, the leading relation, Eq.~(\ref{eq:continuous_slaving}), has a bulk relative $L_2$ error of approximately $10^{-5}$ for these cases. 
The deviation of the fitted exponents from $-1$ and $-2$ is consistent with the exact solution at finite scale separation and does not require an empirical correction to the moment amplitudes.

At fixed relaxation times, $\kappa$ and $\ell_D$ are constant, and increasing $H$ increases the dimensionless width $\kappa H$. 
For the parameters shown in Fig.~\ref{fig:poiseuille_hierarchy}(b),
$\ell_D\simeq129.10$ and $\kappa H$ ranges from $0.387$ to $3.098$.
Thus, the sweep does not remain entirely in the parabolic limit
$\kappa H\ll1$. The $H^{-1}$ and $H^{-2}$ guides describe an intermediate Poiseuille ordering, rather than universal large-$H$ laws. 
For example, in a sufficiently wide channel, the interior approaches $q_\infty$, and Eq.~(\ref{eq:a8_local_exact}) gives $a_{8,\infty}/q_\infty=-2\beta/(15\gamma)$, a finite ratio at fixed rates.

\subsection{Diffuse-wall kinetic structure}
\label{sec:wall_kinetic}

For the diffuse-reflection reconstruction, the wall populations satisfy
\begin{equation}
m_{12,w}=-m_{11,w}
\quad\text{(bottom)},
\qquad
m_{12,w}=+m_{11,w}
\quad\text{(top)}.
\end{equation}
Defining the inward-oriented moment
\begin{equation}
h_w=
\begin{cases}
-m_{11,w}, & \text{bottom},\\
+m_{11,w}, & \text{top},
\end{cases}
\end{equation}
and using $a_8=m_{12}-\eta q_x$ gives the exact wall relation
\begin{equation}
h_w=\eta q_w+a_{8,w}.
\label{eq:wall_exact_closure}
\end{equation}

Extending the leading bulk slaving relation to the wall suggests
\begin{equation}
\ell_s\sim\frac{1}{\gamma},
\end{equation}
with
\begin{equation}
\ell_{s,\mathrm{meas}}
=
\frac{|q_w|}
{|(\partial q_x/\partial n)_w|}.
\end{equation}
The measured wall gradients are evaluated by second-order one-sided differences, for example $g_{2,0}=(-3q_0+4q_1-q_2)/2$ at the bottom wall. 
The corresponding inward derivative is used at the top wall.
The same estimator can be applied to the analytical lattice-node profile in Eq.~(\ref{eq:steady_analytic_profiles}); 
this predicts the reported slip diagnostic without substituting measured wall moments.
This procedure differs from differentiating a continuous interpolation of the profile.

\begin{figure*}[tb]
\centering
\includegraphics[width=0.98\textwidth]{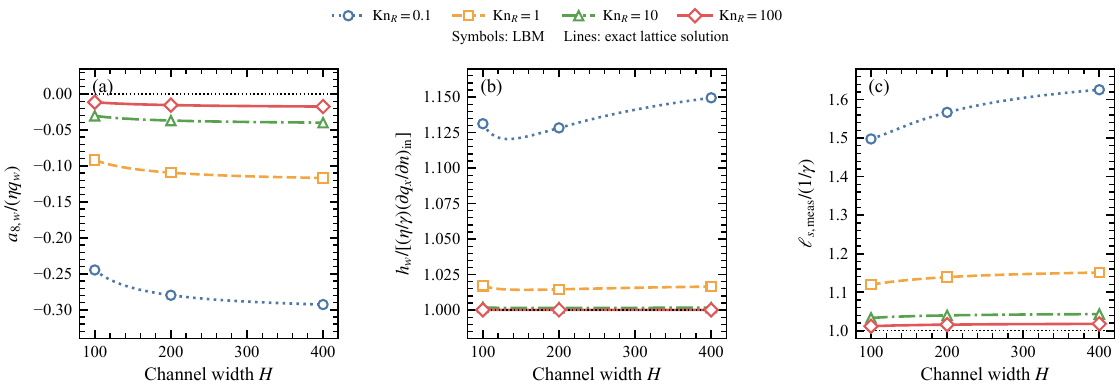}
\caption{
Diffuse-wall kinetic structure at fixed $\mathrm{Kn}_N=0.01$ for
$\mathrm{Kn}_R=0.1$, $1$, $10$, and $100$.
Symbols show the LBM results, and lines show the exact steady lattice
solution without fitted coefficients.
(a) $r_w=a_{8,w}/(\eta q_w)$.
(b) $S_w=h_w/[(\eta/\gamma)(\partial q_x/\partial n)_{\mathrm{in}}]$.
(c) $\ell_{s,\mathrm{meas}}/(1/\gamma)$.
The analytical diagnostics use the same second-order one-sided wall
gradient as the numerical measurements. Each panel shows the average
of the bottom and top wall diagnostics. The reference lines mark
$r_w=0$, $S_w=1$, and $\ell_s/(1/\gamma)=1$, respectively.
}
\label{fig:wall_kinetic}
\end{figure*}

Figure~\ref{fig:wall_kinetic} compares the numerical wall diagnostics (symbols) with the exact steady predictions (lines) at fixed $\mathrm{Kn}_N=0.01$. 
As $\mathrm{Kn}_R$ increases from $0.1$ to $100$, the higher-order ratio $a_{8,w}/(\eta q_w)$ decreases in magnitude, the wall slaving ratio approaches unity, and
$\ell_{s,\mathrm{meas}}/(1/\gamma)$ approaches unity. 
At $\mathrm{Kn}_R=100$, the slaving-ratio deviation is below $2\times10^{-4}$ and the measured slip differs from $1/\gamma$ by only approximately $1$--$2\%$. Thus, the leading wall response is controlled by the
same fast relaxation scale as the transverse bulk hierarchy, whereas
$a_{8,w}$ measures the deviation from lowest-order closure.
More specifically, we define $r_w=a_{8,w}/(\eta q_w)$ and $S_w=h_w/[(\eta/\gamma)g_w]$ using the same inward gradient estimator $g_w$. 
Equation~(\ref{eq:wall_exact_closure}) implies
\begin{equation}
\frac{\ell_{s,\mathrm{meas}}}{1/\gamma}
=\left|\frac{S_w}{1+r_w}\right|.
\label{eq:slip_error_decomposition}
\end{equation}
This relation separates deviations in the wall gradient relation ($S_w\ne1$) from the omission of the higher-order wall amplitude ($r_w\ne0$). 
It holds separately at each wall; Fig.~\ref{fig:wall_kinetic} plots the averages of the corresponding bottom and top diagnostics.
The three panels therefore present related diagnostics rather than independent tests. 
A parameter-only prediction is obtained by evaluating all quantities from the analytical profile, without treating the measured $r_w$ as a fitted boundary correction.

Using the same second-order one-sided wall gradient as in the numerical evaluation, the maximum absolute discrepancies over the twelve cases are $1.20\times10^{-11}$ for $r_w$, $2.96\times10^{-11}$ for $S_w$, and
$4.95\times10^{-11}$ for the normalized slip length. 
The analytical curves reproduce the nonmonotonic width dependence of $S_w$ at $\mathrm{Kn}_R=0.1$. 
Thus, the deviation from the leading slip scale is quantitatively accounted for by the wall kinetic amplitude and the deviation from the leading gradient relation, through the per-wall identity $\ell_s/(1/\gamma)=|S_w/(1+r_w)|$.

Here, the model Knudsen numbers remain fixed, so $\tau_N=0.01H$ and $\tau_R=\mathrm{Kn}_R H$. 
In contrast to Fig.~\ref{fig:poiseuille_hierarchy}(b), both collision rates $\beta$ and $\gamma$ decrease as $H^{-1}$. 
Writing $b=H\beta$ and $g=H\gamma$, $K$ is constant and $\ell_D^2/H^2\to K/(bg)$, while $\kappa H\to\sqrt{bg/K}$.
This is a refinement toward a limiting normalized profile of the specified model, rather than a width sweep at fixed microscopic rates.
It is not exactly a sweep at fixed Guo--Wang physical Knudsen numbers at finite $H$, because of Eq.~(\ref{eq:gw_rate_mapping}).

The nonmonotonic ratio can be resolved further. Let $q'(0)$ denote the derivative of the smooth hyperbolic expression in Eq.~(\ref{eq:steady_analytic_profiles}), and define
\begin{equation}
S_{\rm sm}=\frac{h_0}{(\eta/\gamma)q'(0)}
=\frac{\gamma V}{\eta\kappa},\qquad
F_2=\frac{g_{2,0}}{q'(0)}.
\label{eq:wall_estimator_factors}
\end{equation}
The plotted wall ratio is $S_2=S_{\rm sm}/F_2$, where $S_2$ denotes $S_w$ evaluated using the second-order estimator. 
The exact expression for $F_2$ is given in Appendix~\ref{app:gradient_estimator}.
For $\mathrm{Kn}_R=0.1$, $S_{\rm sm}$ increases monotonically over the
integer widths $100\le H\le400$, whereas $1/F_2$ decreases. At
$H=100,200,400$, $F_2=0.85522,0.97245,0.99372$, respectively.
The decreasing amplification from the gradient underestimate competes
with the increasing smooth-profile ratio, resulting in a minimum of
$S_2$ at the sampled integer width $H=134$.
This minimum is an analytical prediction; the LBM data in the figure
remain the three original widths.
The smooth derivative is an auxiliary diagnostic of the same lattice
solution, not an independently derived continuum boundary law.

In the fixed-model-Knudsen-number limit,
\begin{equation}
F_2\to1,\qquad S_2,S_{\rm sm}\to\frac{K}{\eta}
=1+\frac{2\mathrm{Kn}_N}{\mathrm{Kn}_N+\mathrm{Kn}_R}.
\label{eq:fixed_kn_wall_limit}
\end{equation}
Thus, eliminating the estimator bias need not restore the leading
relation $S_w=1$: at $\mathrm{Kn}_R=0.1$, the limiting value is $1.18182$.
The leading gradient relation also requires strong normal--resistive
scale separation. This distinguishes estimator error from a finite
correction to the moment closure.

\subsection{Scope of the exact solution}
\label{sec:discussion_scope}

The consistency demonstrated in Figs.~\ref{fig:poiseuille_hierarchy} and~\ref{fig:wall_kinetic} verifies the analytical solution of the specified collision--streaming update, energy-drop driving, and diffuse-wall reconstruction.
It establishes the consistency of the numerical implementation with the derived moment closure.
Accuracy relative to the continuous-angle phonon Boltzmann equation additionally depends on the velocity discretization.
For rarefied-gas lattice Boltzmann models, \citet{FeuchterSchleifenbaum2016} demonstrated that accurate bulk quadrature must be accompanied by accurate evaluation of the half-space moments entering diffuse-wall boundary conditions.

The exact wall relation in Eq.~(\ref{eq:wall_exact_closure}) illustrates this dependence on the velocity set.
In the present lattice units, every wall-crossing D2Q8 velocity has $|c_{iy}|=1$.
Together with the symmetry of the diffusely emitted populations, this property yields $m_{12,w}=\mp m_{11,w}$ at the bottom and top walls, respectively.
For a continuous angular distribution, the normal velocity magnitude varies, and the corresponding first and second normal-velocity weights cannot be identified in this way.
The resulting closure therefore characterizes the specified discrete model and boundary treatment.

A complementary comparison with an isotropic gray Callaway Boltzmann equation is provided in the Supplemental Material.
The reference model uses a single group speed and continuous directions on the unit sphere, with finite normal and resistive scattering times and fully diffuse walls.
The speed correspondence introduced in Sec.~\ref{sec:d2q8} matches the coefficients $\chi$ and $\eta$.
To separate velocity-discretization effects from lattice discretization effects, the comparison uses the continuous-space limit of the D2Q8 dynamics and an angularly converged numerical solution of the reference equation.
For the cases examined, the D2Q8 mean heat flux differs from the gray reference by only about $0.4$--$1.3\%$, whereas the corresponding wall heat flux differs by approximately $75$--$94\%$.
The wall amplitude $a_8$ also differs qualitatively between the two velocity representations.
This comparison delineates the physical scope of the exact lattice closure and shows why agreement in an integrated transport observable alone does not establish the accuracy of the wall kinetic structure.

\section{Conclusions}
\label{sec:conclusions}

We derived exact steady solutions linking the heat flux $q$ and the transverse kinetic amplitudes $a_6$ and $a_8$ in a driven D2Q8 Callaway lattice Boltzmann model.
The periodic energy-drop driving and diffuse-wall population reconstruction determine these profiles without fitted parameters.
The resulting scalar difference equation yields an explicit closure for $a_8$ containing both a second-difference term and a driving contribution.
Agreement with the lattice Boltzmann simulations verifies the predicted heat-flux profiles and kinetic-moment amplitudes, not merely their approximate width-scaling exponents.

The same solution predicts the wall response through the exact relation $h_w=\eta q_w+a_{8,w}$.
It separates deviations from the leading slip scale into the higher-order wall amplitude and the deviation from the leading gradient relation.
Evaluating the analytical profile with the same one-sided gradient estimator as the simulations reproduces the observed nonmonotonic width dependence of the wall slaving ratio.
The decomposition of the wall slaving ratio into a smooth-profile ratio and an estimator factor identifies the competing contributions responsible for this behavior.

Fixed-relaxation-time and fixed-model-Knudsen-number width sweeps probe different limits.
The former changes the channel width relative to the transport lengths, whereas the latter refines the lattice toward a limiting normalized profile.
In the latter limit, the estimator bias vanishes, but the wall slaving ratio approaches $K/\eta$, which differs from unity at finite $\beta/\gamma$.
Thus, spatial refinement alone does not recover the leading wall-gradient relation.

The collision eigenspaces provide the organizing structure for these results.
Nested resistive and normal equilibrium projectors separate the D2Q8 population space into three sectors: one conserved energy mode, two heat-flux modes, and five fast kinetic modes.
Streaming and boundary conditions select the moment couplings within this decomposition.
The longitudinal spectral analysis complements the steady channel solution by resolving the finite-wave-number mixing of the conserved and heat-flux sectors into a propagating pair.

These results provide an analytical benchmark for the specified lattice update, driving, and boundary treatment.
Their exactness at the discrete level is distinct from their accuracy as approximations to continuous-angle phonon transport.
The gray Callaway comparison illustrates that close agreement in mean heat flux can coexist with substantial differences in wall observables.
The moment description therefore clarifies both the internal transport mechanism of the discrete model and the need to assess its velocity-space approximation when interpreting wall kinetic structure.

\begin{acknowledgments}
The author thanks Hiroyasu Matsuura, Naoki Takada, Hideaki Maebashi, and Chul-Ho Lee for valuable discussions and comments, and acknowledges support from the Research Institute for Energy Efficient Technologies, AIST.
The author used ChatGPT (OpenAI, GPT-5.6) to assist with reviewing the manuscript and checking analytical derivations.
\end{acknowledgments}

\section*{Data Availability}
The data supporting the findings of this study are openly available in Zenodo at \url{https://doi.org/10.5281/zenodo.23020627}.

\appendix
\section{Derivation of the driven steady solution}
\label{app:steady_closure}

For the $x$-averaged populations, introduce the directional differences
$j_0=f_1-f_3=q-r$, $j_+=f_5-f_6=(r+p)/2$, and
$j_-=f_8-f_7=(r-p)/2$. Their steady interior streaming relations are
\begin{align}
j_0(y)&=j_0^*(y)+4G/9,\nonumber\\
j_+(y)&=j_+^*(y-1)+G/18,\nonumber\\
j_-(y)&=j_-^*(y+1)+G/18.
\label{eq:directional_steady_streaming}
\end{align}
The first equation also holds at the walls and gives
Eq.~(\ref{eq:a8_local_exact}). The other two equations couple the
transverse planes.

Let $\mathcal S_y f_y=(f_{y+1}+f_{y-1})/2$ and
$\delta_y f_y=(f_{y+1}-f_{y-1})/2$, so that
$\mathcal S_y=I+\Delta_2/2$ and
$\delta_y^2=\mathcal S_y^2-I$. Summing and subtracting the last two
relations in Eq.~(\ref{eq:directional_steady_streaming}), and using
$r=Kq-D_gG$ and $r^*=Lq-\sigma D_gG$, gives
\begin{align}
(K-L\mathcal S_y)q&=\chi G-\sigma\delta_y p,\nonumber\\
(I-\sigma\mathcal S_y)p&=-L\delta_y q.
\label{eq:steady_coupled_operators}
\end{align}
Eliminating $p$ cancels the $\mathcal S_y^2$ terms and yields
\begin{equation}
\beta\gamma q-
\frac{K(2-\gamma)-\beta}{2}\Delta_2q=\gamma\chi G,
\end{equation}
which is Eq.~(\ref{eq:steady_scalar_recurrence}). These are interior
relations; boundary values are fixed by the population reconstruction.

At each wall, the outgoing populations are proportional to $w_i$, with
the proportionality constant chosen so that their sum equals the
incoming sum. Since all wall-crossing velocities have $|c_{iy}|=1$,
this imposes zero normal heat flux. Equal outgoing diagonal weights
give $j_+(0)=0$ and $j_-(H)=0$, or $p_0=-r_0$ and $p_H=r_H$.

Substituting a spatial mode $q_y-q_\infty\propto e^{\kappa y}$ in
Eq.~(\ref{eq:steady_scalar_recurrence}) gives
$2\ell_D^2(\cosh\kappa-1)=1$. The second relation in
Eq.~(\ref{eq:steady_coupled_operators}) then gives the coefficient
$V$ in Eq.~(\ref{eq:steady_spatial_rate}). Reflection symmetry combines
the spatial modes into the even heat-flux and odd shear-moment
profiles in Eq.~(\ref{eq:steady_analytic_profiles}). Applying
$p_0=-r_0$ fixes $A_c$ as in Eq.~(\ref{eq:steady_amplitude});
the top-wall relation follows by symmetry. Direct substitution in
Eq.~(\ref{eq:directional_steady_streaming}) verifies the interior
update and the reconstructed wall constraints.

For Fig.~\ref{fig:poiseuille_hierarchy}, the analytical moments are
sampled at the same lattice nodes as the numerical data. Their own
centerline or full-domain maximum heat flux supplies the normalization,
and their bulk RMS values use the same exclusion widths. At a wall,
the second-order difference applied to
Eq.~(\ref{eq:steady_analytic_profiles}) predicts the slip diagnostic
used in Fig.~\ref{fig:wall_kinetic}. 
The solution is exact for the specified steady-state lattice equations and boundary reconstruction, but is not an exact solution of the continuous phonon Boltzmann equation.

\section{Wall-gradient estimator factor}
\label{app:gradient_estimator}

Let $t=\kappa H/2$. Applying the second-order one-sided difference to
$q_y=q_\infty+A_c\cosh[\kappa(y-H/2)]$ and dividing by
$q'(0)=-\kappa A_c\sinh t$ gives
\begin{equation}
F_2=\frac{4\sinh\kappa-\sinh(2\kappa)}{2\kappa}+\frac{\coth t}{2\kappa}
\bigl[3-4\cosh\kappa+\cosh(2\kappa)\bigr].
\label{eq:gradient_factor_exact}
\end{equation}
Reflection symmetry gives the same factor at the top wall. 
At fixed model Knudsen numbers, $t$ has a finite positive limit and
$\kappa=O(H^{-1})$; hence,
$F_2=1-\kappa^2/3+(\kappa^3/4)\coth t+O(\kappa^4)$.
Moreover, $1-\sigma\cosh\kappa=\gamma+O(H^{-2})$; hence,
$V=K\kappa/\gamma+O(H^{-1})$ and
$S_{\rm sm}\to K/\eta$. These results give
Eq.~(\ref{eq:fixed_kn_wall_limit}).

\bibliography{refs}

\end{document}